\newcommand{\diracslash}[1]{#1\llap{/\kern2pt}}

\def\bearr{\begin{eqnarray}}

\def\eearr{\end{eqnarray}}

\newcommand{\be}{\begin{equation}}

\newcommand{\ee}{\end{equation}}

\newcommand{\bea}{\begin{eqnarray}}

\newcommand{\eea}{\end{eqnarray}}

\newcommand{\ba}[1]{\begin{array}{#1}}

\newcommand{\ea}{\end{array}}

\newcommand{\eqrf}[1]{Eq.\ (\ref{#1})}

\newcommand{\eqrftw}[2]{Eqs.\ (\ref{#1}) and (\ref{#2})}

\documentclass[preprint,secnumarabic,floats,amssymb,bibnotes,nofootinbib,showpacs,tightenlines]{revtex4}

\usepackage{graphicx}
\usepackage{fixltx2e}
\usepackage{amsmath}

\usepackage{latexsym}

\usepackage{epsfig}

\usepackage{subfigure}

\begin{document}
\title{Constraints on electromagnetic form factors of sub-GeV dark matter from the Cosmic Microwave Background anisotropy}
\author{Gaetano Lambiase $^a$, Subhendra Mohanty$^b$, Akhilesh Nautiyal$^c$ and Soumya Rao$^b$}
\affiliation{$^a$ Dipartimento di Fisica “E.R Caianiello”, Universit degli Studi di Salerno, Via Giovanni Paolo II, 132 - 84084 Fisciano (SA), Italy \\ $^b$ Theory Division, Physical Research Laboratory,
Navrangpura, Ahmedabad 380 009, India\\
$^c$ Department of Physics, Malaviya National Institute of Technology, JLN Marg, Jaipur 302017,India }

\begin{abstract}
We consider dark matter which have non-zero electromagnetic form factors like electric/magnetic dipole moments and anapole moment for fermionic dark matter and Rayleigh form  factor for scalar dark matter. We consider dark matter mass $m_\chi > \cal{ O}({\rm MeV})$ and put constraints on their mass and electromagnetic couplings from CMB and LSS observations. Fermionic dark matter with non-zero electromagnetic form factors  can annihilate to $e^+ e^-$  and scalar dark matter can annihilate to $2\gamma$ at the time of recombination and distort the CMB. 
We analyze dark matter with multipole moments with Planck  and BAO  observations.
We find upper bounds on anapole moment $g_{A}<7.163\times 10^{3} \text{GeV\textsuperscript{-2}}$,
electric dipole moment ${\cal D}<7.978\times 10^{-9} \text{e-cm}$, magnetic dipole moment
${\mu}<2.959\times 10^{-7} \mu_B$,  and the bound on  Rayleigh form factor of dark 
matter  is $g_4/\Lambda_4^2<1.085\times 10^{-2}\text{GeV\textsuperscript{-2}}$ with $95\%$C.L.
\end{abstract}

\maketitle
\section{Introduction}
It is well accepted that formation of large scale structures and the rotation curves of galaxies 
require an extra dark matter component beyond than the known particles of the standard model. 
The particle properties of this  dark matter are, however, completely unknown. 
Direct detection experiments, which rely on nuclear scattering, have ruled out a large 
parameter space. But, these techniques  are not efficient in measuring dark matter of sub-GeV mass \cite{Undagoitia:2015gya, Mohanty:2020pfa}. To measure sub-GeV mass dark matter a suitable method  
is scattering electrons from heavy 
atoms \cite{Essig:2011nj,Essig:2012yx, Agnes:2018oej, Aprile:2019xxb, Catena:2019gfa, Catena:2020tbv}. 
Dark matter with non-zero electric or magnetic  dipole moments \cite{Pospelov:2000bq, Sigurdson:2004zp, Masso:2009mu} or anapole moment \cite{ Ho:2012bg, DelNobile:2014eta} can be very 
effective in scattering electrons. The electromagnetic form factors can be viewed as effective 
operators \cite{Primulando:2015lfa, DeSimone:2016fbz, Chu:2018qrm, kavanagh}, which arise by integrating out the 
heavy particles in a ultraviolet complete theory \cite{Kopp:2014tsa}.

The electromagnetic couplings of dark matter can be constrained from cosmic microwave background 
and large scale structures observations. 
The electric and magnetic dipole moment vertex can give rise to dark matter-baryon coupling.  For heavy dark matter ($\sim100\, {\rm GeV}$)
the baryon drag on the dark matter will show up in structure formation and CMB \cite{Sigurdson:2004zp}. Light dark matter ($\cal{O}({\rm MeV})$ will annihilate to radiation and lower the effective neutrino number ($N_{\rm eff}$) \cite{Ho:2012br,Boehm:2013jpa,Brust:2013ova}. 

In this paper we will analyze the effect of light dark matter with electromagnetic form factors 
on  CMB anisotropy and polarization from dark matter annihilation to $e^+ e^-$ or photons close 
to recombination era, $z\sim 1100$. The effects of annihilating dark matter on CMB are studied in
\cite{Chen:2003gz,Padmanabhan:2005es,Galli:2009zc,Slatyer:2009yq,Finkbeiner:2011dx}. 
Production of relativistic $e^+, e^-$  heats up the thermal gas and ionizes the neutral
Hydrogen, which increases the free electron fraction. Due to this increased free electron
fraction there is a broadening of the last scattering surface and suppression of CMB temperature
anisotropy. The low-$l$ correlations between polarization fluctuations are also enhanced
due to increased freeze-out value of the ionization fraction of the universe after recombination. 
These effects on CMB are significant and can be used to put constraints on 
thermal averaged annihilation cross-section $\langle \sigma v \rangle$. Planck-2018 reports
 $\langle \sigma v \rangle < \left(3.2\times 10^{-28}/f\right)\times\left(M_{DM}/
\left(\text{GeV}/\text{c\textsuperscript{2}}\right)\right)\text{cm\textsuperscript{3}/\text{s}}$
 for velocity independent thermal  average cross-section \cite{Aghanim:2018eyx}.  Here $f$ is 
the fraction of energy injected to the intergalactic medium (IGM)  by annihilating dark matter.
Forecasts for upcoming CMB experiments such as AdvACTPol, AliCPT, CLASS, Simons Array, Simons
Observatory, and SPT-3G has been studied by \cite{Cang:2020exa} in detecting 
decaying/annihilating dark matter, and it is found that 
$\langle \sigma v \rangle \sim \left( 10^{-29}/f\right)\times\left(M_{DM}/
\left(\text{GeV}/\text{c\textsuperscript{2}}\right)\right)\text{cm\textsuperscript{3}/\text{s}}$.   

The annihilation $\chi \chi \rightarrow e^+ e^-$ occur with one dipole or anapole vertex and 
the annihilation cross sections are quadratic in dipole or anapole moments. For fermionic dark 
matter the $\chi \chi \rightarrow \gamma \gamma$ annihilation cross section are quartic in 
dipole moments and the bounds from this process are much weaker \cite{Arellano-Celiz:2019pax} 
 than the ones we derive in this paper from CMB. 
For anapole dark matter the cross section for the process $\chi \chi \rightarrow \gamma \gamma$ 
is zero \cite{Ho:2012bg}. Scalar dark matter can have dimension-6  Rayleigh operator vertex with two-photons. For such Rayleigh dark matter the leading order contribution to annihilation will be from the $\phi \phi \rightarrow \gamma \gamma$ process which can distort the CMB near recombination and from this we put bounds on the dark-matter photon Rayleigh coupling. 

This paper is organized as follows. In Section \ref{Electromagnetic} we list the 
electromagnetic form factors of dark matter which we shall constrain from CMB data. 
In Section \ref{CMB} we discuss the physics of recombination and the effect of dark matter 
annihilation on the CMB. In Section \ref{Results} we compare the CMB analysis with data from 
Planck and BAO and using  COSMOMC we put constraints on the dark matter form factors. 
In Section \ref{Comparison} we compare our bounds with earlier results and from other experiments,
 and in Section \ref{Conclusions} we summarize our results and give our 
conclusions.

\section{Electromagnetic form factors of dark matter}\label{Electromagnetic}
Spin-1/2 dark matter can have the following electromagnetic form factors. 
These are the magnetic moment described by the dimension-5 operators,
\be
{\cal L}_{\rm magnetic}=\frac{g_1}{\Lambda_1} \bar \chi \sigma^{\mu \nu} \chi F_{\mu \nu}
\label{magnetic}
\ee
where $g_1$ is a dimensionless coupling and $\Lambda_1$ is the mass scale of the particles in the loop which generate the dipole moment. The magnetic moment of Dirac fermions is $\mu=2 g_1/\Lambda_1$ and the operator (\ref{magnetic}) is zero for Majorana fermions.

Similarly electric dipole operator is of dimension-5,
\be
{\cal L}_{\rm electric}=\frac{g_2}{\Lambda_2} \,i\, \bar \chi \sigma^{\mu \nu} \gamma_5 \chi F_{\mu \nu}
\label{electric}
\ee
where the electric dipole moment of Dirac fermions is ${\cal D}= 2g_2 /\Lambda_2$ and the operator (\ref{electric}) is zero for Majorana fermions.

Finally the anapole moment is a dimension-6 operator
\be
{\cal L}_{\rm anapole}=\frac{g_3}{\Lambda_3^2} \,i\, \bar \chi \gamma^{\mu} \gamma_5 \chi \partial^\nu \,F_{\mu \nu}
\label{anapole}
\ee
This operator is non-zero for Dirac as well as Majorana  fermions and the coefficient $g_A= g_3/( \Lambda_3^2)$ is called the anapole moment of $\chi$. 

Stringent bounds on sub-GeV mass dark matter are put from the observation of $\chi e^-\rightarrow \chi e^-$ scattering \cite{Essig:2011nj} in direct detection experiments like Xenon-10 \cite{Essig:2012yx},  DarkSide \cite{Agnes:2018oej} and Xenon-1T \cite{Aprile:2019xxb}.

Using the experimental limits on dark matter electron scattering from Xenon-10, Xenon-1T and DarkSide bound on the electric dipole, magnetic dipole and anapole form factors of dark matter have been put in ref.\cite{Catena:2019gfa, Catena:2020tbv}.

Real and complex scalar dark matter can have interaction with 2-photons by dimension-6 Rayleigh operator
\be
{\cal L}_{2\phi\, 2\gamma} = \frac{g_4}{\Lambda_4^2} \phi^* \phi F_{\mu \nu} F^{\mu \nu} 
\label{Rayleigh}
\ee
These will contribute to $\phi \phi \rightarrow \gamma \gamma$ annihilations which can be constrained from CMB \cite{kavanagh}. In the absence of CP violation the $\tilde F F$ operator does not arise. The annihilation $\phi \phi \rightarrow \gamma \gamma$ takes place via s-wave in the leading order and  cross section  
$\sigma( \phi \phi \rightarrow \gamma \gamma) v \simeq (g_4^2) m_\phi^2 /\Lambda_4^4$ \cite{kavanagh}. 
Bounds on the operator  (\ref{Rayleigh}) from Xenon1T  \cite{Aprile:2019xxb} and gamma ray 
searches  from dwarf spheroidal satellites (dSphs) \cite{Fermi-LAT:2016uux} and halo of the 
Milky way \cite{Ackermann:2015lka}  by Fermi-LAT  are 
obtained in ref. \cite{kavanagh} for dark matter with mass larger than $\cal{O}(\rm GeV)$.

\section{Thermal history of the universe with annihilating dark matter}\label{CMB}
Recombination occurs around $z=1100$ when electrons and protons combine together
to form neutral hydrogen. If the annihilation cross-section of dark matter 
particles is sufficiently large, it can modify the history of recombination and
hence can leave a clear imprint on CMB power spectrum.
The shower of particles produced due to annihilation can interact with the thermal
gas in three different ways. ({\it i}) The annihilation products can ionize the 
thermal gas, ({\it ii}) can induce induce Lyman-$\alpha$ excitation of the hydrogen
that will cause more electrons in $n=2$ state and hence increase the ionization 
rate and ({\it iii}) can heat the plasma. Due to the first two effects the evolution
of free electron fraction $\chi_e$ changes and the last effect changes the 
temperature of baryons. The equation governing the evolution of ionization
fraction in the presence of annihilating particles is given as
\be
\frac{d\chi_e}{dt} = \frac{1}{\left(1+z\right)H(z)}\left[R_s(z)-I_s(z)-I_X(z)\right].
\label{dxedt1}
\ee
Here $R_s$ is the standard recombination rate, $I_s$ is the ionization rate
due to standard sources and $I_X$ is the ionization rate due to annihilating dark
matter particles. The computation of standard recombination rate was done in 
\cite{Peebles:1968ja,Zeldovich:1969en,Seager:1999km} and it is described as
\be
\left[R_s(z)-I_s(z)\right] = C\times\left[\chi_e^2n_H\alpha_B-\beta_B\left(1-\chi_e\right)
e^{-h_p\nu_{2s}/k_BT_b}\right].\label{stdrec}
\ee
Here $n_H$ is the number density of hydrogen nuclei, $\alpha_B$ and $\beta_B$ are the effective
recombination and photo-ionization rates for principle quantum numbers $\ge 2$ in Case B 
recombination,$\nu_{2s}$ is the frequency of the $2s$ level from the ground state 
and $T_b$ is the temperature of 
the baryon gas. The factor $C$ appearing in Eqn.~(\ref{stdrec}) is given by:
\be
C=\frac{\left[1+K\Lambda_{2s1s}n_H\left(1-\chi_e\right)\right]}
{\left[1+K\Lambda_{2s1s}n_H\left(1-\chi_e\right)+K\beta_Bn_H\left(1-\chi_e\right)\right]}.
\ee
Here $\Lambda_{1s2s}$ is the decay rate of the metastable $2s$ level, $n_H\left(1-\chi_e\right)$
is the number of neutral ground state $H$ atoms and $K= \frac{\lambda_\alpha^3}{8\pi H(z)}$, where
$H(z)$ is the Hubble expansion rate at redshift $z$ and $\lambda_\alpha$ is the wavelength of the
$Ly-\alpha$ transition from the $2p$ level to the $1s$ level.

The term $I_X$ appearing in \eqrf{dxedt1} represents the evolution of free electron density 
due to nonstandard sources. In our case it is due to annihilation of dark matter during recombination.
which increases the ionization rate in two ways.  ({\it i}) By direction ionization from the ground
state and ({\it ii}) by additional $Ly-\alpha$ photons, which boosts the population at $n=2$ 
increasing the  the rate of  photoionization by CMB. Hence, the ionization rate $I_X$ due to dark
matter annihilation is expressed as 
\be
I_{X}(z) = I_{Xi}(z)+I_{X\alpha}(z). \label{ionrate}
\ee
Here $I_{Xi}(z)$ represents the ionization rate due to ionizing photons and $I_{X\alpha}$ represents
the ionization rate due to $Ly-\alpha$ photons.

The rate of energy release $\frac{dE}{dt}$ per unit volume by a relic 
self-annihilating dark matter
particle can be expressed in terms of its thermally averaged annihilation cross-section 
$\langle \sigma v \rangle$ and mass $m_\chi$ as
\be
\frac{dE}{dt} = 2g\rho_c^2 c^2\Omega_{DM}^2\left(1+z\right)^6 f(z) \frac{\langle \sigma v \rangle}{m_\chi},
\label{dedt}
\ee 
where $\Omega_{DM}$ is the dark matter density parameter, $\rho_c$ is the critical density 
today, $g$ is degeneracy factor $1/2$ for Majorana fermions and $1/4$ for Dirac fermions,
and $f(z)$ is the fraction of energy absorbed by the CMB plasma, 
which is $O(1)$ factor and depends on redshift. A detailed calculation of redshift
dependence of $f(z)$ for various annihilation channels is done in 
\cite{Slatyer:2009yq,Huetsi:2009ex,Evoli:2012qh,Galli:2013dna,Madhavacheril:2013cna} using
generalized parameterizations or principle components. 
It is shown in \cite{Galli:2011rz,Giesen:2012rp,Finkbeiner:2011dx}
 that the redshift dependence
of $f(z)$ can be ignored up to a first approximation, since current CMB data are sensitive to 
energy injection over a relatively narrow range of redshift, typically $z \sim 1000-600$. Hence
$f(z)$ can be replace with a constant $f$, which we take as $1$ for our analysis. Here we use
'on-the-spot' approximation, which assumes that the  energy released due to dark matter
annihilation is absorbed by IGM locally \cite{Zhang:2006fr,Zhang:2007zzh,Galli:2009zc}.

The terms appearing on the right hand side of \eqrf{ionrate} are related to the rate of energy
release  as
\bea
I_{Xi} &=& C \chi_i \frac{\left[dE/dt\right]}{n_H(z)E_i}\label{ichii}\\
I_{X\alpha} &=& \left(1-C\right)\chi_\alpha\frac{\left[dE/dt\right]}{n_H(z)E_\alpha}\label{ichialpha}.
\eea 
Here $E_i$ is the average ionization energy per baryon, $E_\alpha$ is the difference  in binding
energy between the $1s$ and $2p$ energy levels of a hydrogen atom, $n_H$ is the number density
of Hydrogen Nuclei, and $\chi_i$ and $\chi_\alpha$  represent the fraction of 
energy going ionization and $Ly-\alpha$ photons respectively; which can be expressed in terms 
of free electron fraction as $\chi_i=\chi_\alpha=\left(1-\chi_e\right)/3$ \cite{Chen:2003gz}.

A fraction of energy released by annihilating dark matter particles also goes into heating the 
baryon gas, which modifies the evolution equation for the matter temperature $T_b$ by contributing
one extra term $K_h$ as
\be
\left(1+z\right)\frac{dT_b}{dz} = \frac{8\sigma_T a_R T_{CMB}^4}{3m_e c H(z)}
\frac{\chi_e}{1+f_{He}+\chi_e}\left(T_b-T_{CMB}\right) -\frac{2}{3k_B H(z)}\frac{k_h}
{1+f_{He}+\chi_e} +2 T_b \label{mattert}.
\ee
Here the nonstandard term $K_h$ arising due to annihilating dark matter is given in terms of rate
of energy release as
\be
K_h = \chi_h\frac{\left(dE/dt\right)}{n_H(z)},
\ee
with $\chi_h= \left(1+2\chi_e\right)/3$ being the fraction of energy going into heat.

In this work we consider annihilating dark matter with electromagnetic form factors. 
We will now obtain the energy deposition rate for dark matter with anapole moment, electric dipole
moment and magnetic dipole moment. One can define a quantity
$p_{ann}$ that depends on the properties of dark matter particles as
\be
p_{ann} =f \frac{\langle \sigma v \rangle}{m_\chi}. \label{pann}
\ee  
The current constraint on $p_{ann}$ with velocity independent $\langle \sigma v \rangle$ is
$1.795\times 10^{-7}$ m\textsuperscript{3}s\textsuperscript{-1}kg\textsuperscript{-1} 
$95\%\;C.L$ from Planck-2018 \cite{Aghanim:2018eyx}. In our analysis we
will use various electromagnetic form factors and mass of the dark matter 
as our model parameters rather than $p_{ann}$. Hence we will express energy
deposition rate in terms of these parameters for annihilating dark matter with
anapole and dipole moments.
\subsection{ Dark matter with anapole moment}
The annihilation cross-section for dark matter with anapole moment is given as \cite{Ho:2012bg},
\be
\langle \sigma v \rangle_{\chi \chi \rightarrow e^+ e^-}  = \frac{2g_A^2 \alpha m_\chi^2}{3}
v_{rel}^2.\label{sigmaann}
\ee
where $\alpha=e^2/(4 \pi)$ and $v_{rel}$ is average relative velocity between the annihilating 
dark matter particles in the centre of mass frame. The thermally averaged velocity can 
be expressed in terms of temperature  by $\frac12\left(\frac12m_\chi\right)
\langle v^2_{rel}\rangle = \frac32 T$. Hence the cross-section (\ref{sigmaann}) can be
expressed in terms of temperature as
\be
\langle \sigma v \rangle_{\chi \chi \rightarrow e^+ e^-}  
=4g_A^2 \alpha m_\chi^2\left(\frac{T}{m_\chi}\right).\label{sigmaannT}
\ee
After decoupling the temperature of the  dark matter behaves as $T\propto \left(1+z\right)^2$.
Assuming the decoupling temperature of the dark matter $T_d$ of the order 
of $\frac{m_\chi}{10}$ we get
\bea
T&=&T_d\frac{\left(1+z\right)^2}{\left(1+z_d\right)^2}=T_d\frac{T_0^2}{T_{\gamma d}^2}
\left(1+z\right)^2\nonumber\\
&=&\frac{10 T_0^2}{m_\chi}\left(1+z\right)^2.\label{Tdm}
\eea
Here  $z_d$ and $T_{\gamma d}$ are the redshift of dark matter decoupling and the temperature
of photons at that redshift respectively, which is same as $T_d$. $T_0$ is the current temperature
of CMB. Using \eqrf{Tdm} the annihilation cross-section (\ref{sigmaannT})  becomes
\be
\langle \sigma v \rangle_{\chi \chi \rightarrow e^+ e^-}  
=40g_A^2 \alpha T_0^2\left(1+z\right)^2.\label{sigmaannrs}
\ee
Hence using \eqrftw{sigmaannrs}{pann} we can obtain the expression for $p_{ann}$ as
\be
p_{ann} =\frac{40g_A^2 \alpha T_0^2}{m_\chi}\left(1+z\right)^2.\label{panniapm}\\
\ee
As mentioned earlier we will choose $f\sim 1$ here. 
Since $p_{ann}$ is velocity dependent, the rate of energy release given by \eqrf{dedt} will be
\be
\frac{dE}{dt} = \rho_c^2 c^2\Omega_{DM}^2\frac{40g_A^2 \alpha T_0^2}{m_\chi}\left(1+z\right)^8
\label{dedtapm}.
\ee
Here the redshift dependence of the energy deposition rate is modified as 
compared to (\ref{dedt}).

\subsection{Dark matter with electric dipole moment}
For DM with electric dipole moment the annihilation cross-section is given by \cite{Masso:2009mu},
\be
\langle \sigma v \rangle_{\chi \chi \rightarrow e^+ e^-} = \frac{\alpha \mathcal{D}^2}{12}
v^2_{rel}
\label{Dann}
\ee
where $v_{rel}$ is the relative velocity of two annihilating WIMPS. For thermal averaged 
cross-section $T=m_\chi\langle v^2_{rel}\rangle/3$. So the annihilation cross-section for
dark matter can be expressed in terms of temperature as
\be
\langle \sigma v \rangle_{\chi \chi \rightarrow e^+ e^-} = 
\frac{\alpha \mathcal{D}^2}{4}\left(\frac{T}{m_\chi}\right).
\ee
Assuming $T_d\sim \frac{m_\chi}{10}$ and  using \eqrf{Tdm} for temperature of the dark matter 
the annihilation cross-section for dark matter with electric dipole moment becomes
\be
\langle \sigma v \rangle_{\chi \chi \rightarrow e^+ e^-} =
\frac{5\alpha \mathcal{D}^2T_0^2}{2m_\chi^2}\left(1+z\right)^2.\label{sigmaedmrs}
\ee
Hence using (\ref{pann}) we get
\be
p_{ann} = \frac{5\alpha \mathcal{D}^2T_0^2}{2m_\chi^3}\left(1+z\right)^2.\label{panniedm}
\ee

 In this case  the energy deposition rate will be
\be
\frac{dE}{dt} = \frac12\rho_c^2 c^2\Omega_{DM}^2\frac{5\alpha \mathcal{D}^2T_0^2}{2m_\chi^3}
\left(1+z\right)^8
\label{dedtedm}.
\ee
 Here also the redshift dependence is modified as compared to (\ref{dedt}), since the thermally
averaged cross-section is velocity dependent.

\subsection{Dark matter with magnetic dipole moment}
For dark matter with magnetic dipole moment the annihilation cross-section is given as \cite{Masso:2009mu},
\be
\langle \sigma v \rangle_{\chi \chi \rightarrow e^+ e^-}  = \alpha \mu^2,
\label{magann}
\ee
and hence
\be
p_{ann} = \frac{\alpha \mu^2}{m_\chi}.\label{pannimdm}
\ee
Here the annihilation cross-section does not depend on the velocity of dark matter so the 
energy deposition rate will be
\be
\frac{dE}{dt} = \frac12\rho_c^2 c^2\Omega_{DM}^2\frac{\alpha \mu^2}{m_\chi}
\left(1+z\right)^6,
\label{dedtmdm}.
\ee 
which has the same redshift dependence as in \eqrf{dedt}. 

\subsection{Rayleigh dark matter}
For scalar dark matter with Rayleigh coupling (\ref{Rayleigh}) the annihilation cross-section is given by \cite{kavanagh},
\be
\langle \sigma v \rangle_{\phi \phi \rightarrow \gamma \gamma}  = \frac{(g_4^2) m_\phi^2}{ \Lambda_4^4}  ,
\ee
and hence
\be
p_{ann} = \frac{(g_4^2) m_\phi}{ \Lambda_4^4}.\label{pannimdm}
\ee
Here again the annihilation cross-section is independent of the velocity of dark matter, so the 
energy deposition rate will be same as (\ref{dedt}). 
\be
\frac{dE}{dt} = \frac12\rho_c^2 c^2\Omega_{DM}^2 \frac{(g_4^2) m_\phi}{ \Lambda_4^4}
\left(1+z\right)^6
\label{dedtrdm}.
\ee

\section{CMB  constraints on various multipole moments of dark matter} \label{Results}
As mentioned earlier annihilating dark matter increases the ionization fraction during
recombination and heats the plasma. 
Hence the evolution equations of free electron fraction
and matter temperature  get modified as given by \eqrf{dxedt1} and \eqrf{mattert} respectively.
The non-standard ionization rate $I_X$ to compute free electron fraction 
can be obtained using \eqrf{ionrate} along with \eqrftw{ichii}{ichialpha}.
We use these equations along with energy deposition rates (\ref{dedtapm}), (\ref{dedtedm}), 
(\ref{dedtmdm}) and (\ref{dedtrdm}) for dark matter with anapole moment, electric dipole moment 
and magnetic dipole moment,
and Rayleigh coupling to modify RECFAST routine \cite{Seager:1999km}
  in CAMB \cite{Lewis:1999bs}. We have also checked our analysis using 
CosmoRec \cite{Chluba:2010ca}   and HyRec \cite{AliHaimoud:2010dx, Giesen:2012rp}  code
instead of RECFAST, and we found similar results. With this we obtain modified 
 theoretical angular power spectra, 
which can be used to compute the bounds on various 
electromagnetic form factors and mass of the dark matter from 
Planck-2018 data using
COSMOMC \cite{Lewis:2002ah}. 
The priors for the multipole moments and mass of the dark matter are given in 
Table~\ref{tab:priors}. All these priors are sampled logarithmically to cover a larger range
for the new parameters. We also vary the other six parameters of $\Lambda$CDM model with 
priors given in \cite{Ade:2013zuv}. We have imposed flat priors for all parameters.

\begin{table}[h]
\begin{tabular}{l c}
Type of dark matter coupling  & Priors\\
\hline\hline
\\
Anapole             & \begin{tabular}{@{}ll@{}}
$5.0 < ${\boldmath $\ln\left(10^{9}\left(g_{A}/GeV^{-2}\right)\right)$}$ <40$ &
$\;\;-3.0 < ${\boldmath$\log_{10}\left(m_{\chi}/GeV\right)$} $<2.0$\\
\end{tabular}\\
Electric dipole     & \begin{tabular}{@{}ll@{}}
$-5.0<$ {\boldmath$\ln(10^{18}({\cal D}/(e-cm))$} $< 40$ &
$\;\;\;-3.0 <$ {\boldmath$\log_{10}\left(m_{\chi}/GeV\right)$} $<2.0$\\
\end{tabular}\\
Magnetic dipole     & \begin{tabular}{@{}ll@{}}
$-10.0<$ {\boldmath$\ln(10^{9}(\mu/\mu_B))$} $< 15.0$
$\;\;\;\;-3.0 <$ {\boldmath$\log_{10}\left(m_{\chi}/GeV\right)$} $<2.0$\\
\end{tabular}\\
Rayleigh coupling & \begin{tabular}{@{}ll@{}}
$-10.0<$ {\boldmath$\ln(10^{9}g_{4}/(\Lambda_{4}^{2} GeV^{-2}))$} $<20.0$
$\;\;\;\;-12.0$ {\boldmath$<\log_{10}\left(m_{\chi}/GeV\right)$} $<2.0$\\
\\
\end{tabular}\\
\hline\hline  
\end{tabular}
\caption{Priors on input parameters for annihilating dark matter.}
\label{tab:priors}
\end{table}

We use the lower bound for the mass of dark matter with anapole moment, and electric and magnetic
dipole moment as $1$ MeV since the annihilation channel for this case is 
$\chi\; \chi \rightarrow\; e^+\,e^-$. However, in case of scalar dark matter with 
Rayleigh coupling, \cite{kavanagh},  
the dark matter annihilates to photons having energy around $1$ eV 
during recombination, which is used as the lower bound for mass of Rayleigh dark matter. 
We also use BAO and Pantheon data along with Planck-2018 observations for our analysis. We 
perform MCMC convergence diagnostic tests on 4 chains using the Gelman and Rubin 
"variance of mean"/"mean of chain variance" R-1 statistics for each parameter.

\begin{figure}[h]
\begin{center}
\subfigure[Marginalized constraints]{
 \includegraphics[width=7cm, height = 6cm]{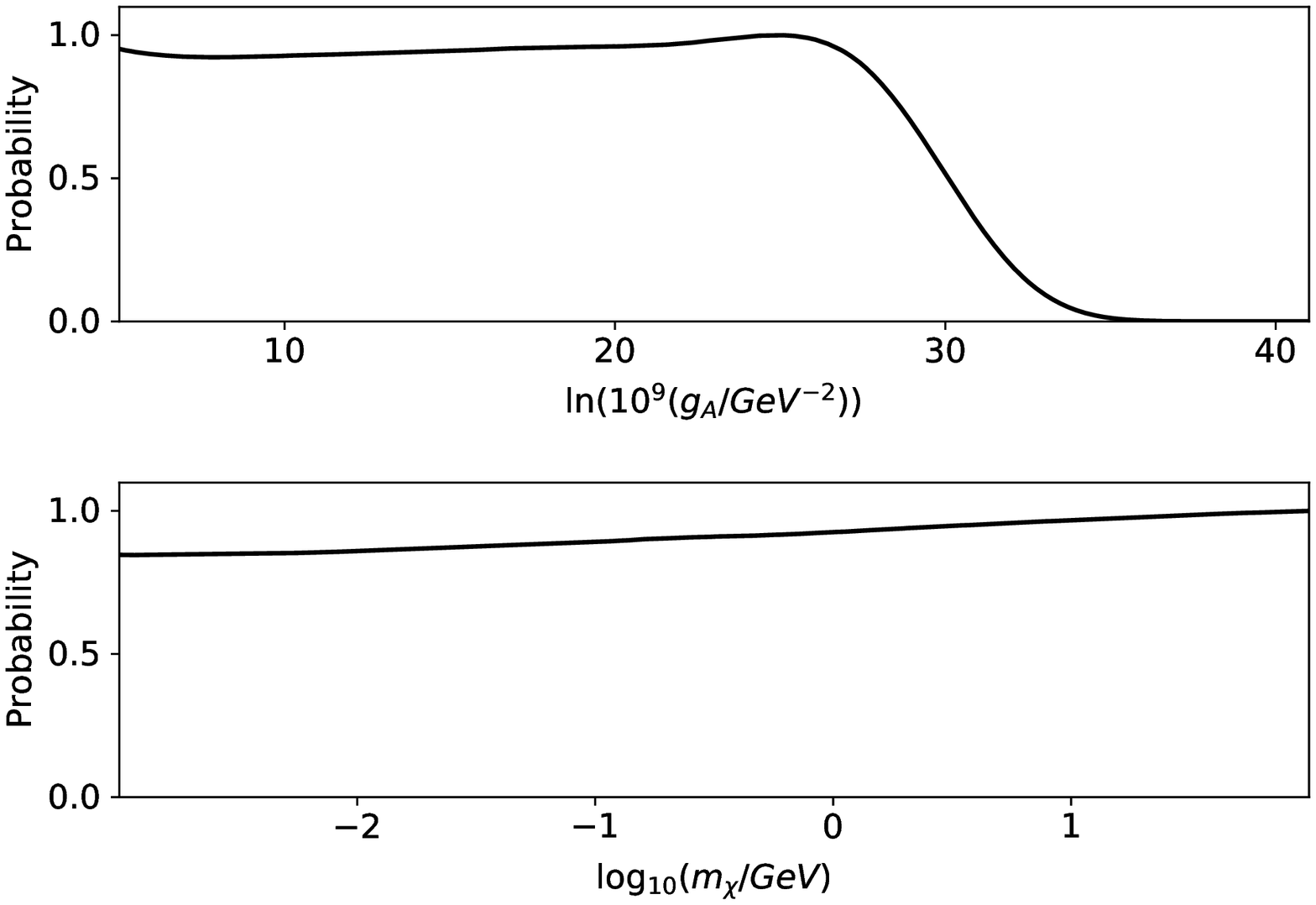}
}
\subfigure[ Joint 68\% CL and 95\%CL constraints]{
 \includegraphics[width=7cm, height = 6cm]{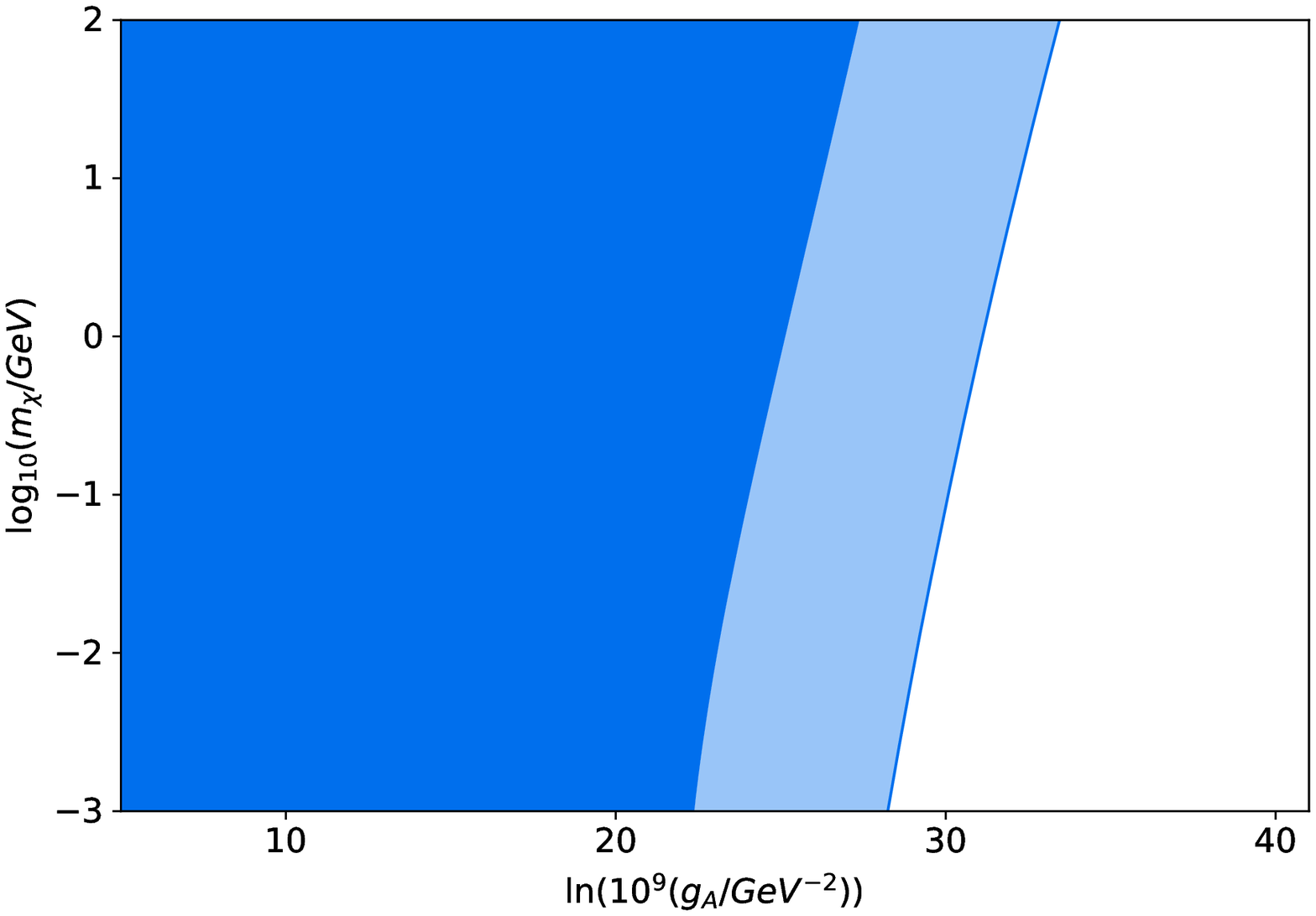}
}
\caption{Constraints for  anapole moment and mass of the dark matter using Planck-2018 and BAO data}
\label{fig:apmconstr}
\end{center}
\end{figure}

The constraints obtained for anapole moment and mass of the dark matter along
with other six parameters of $\Lambda$CDM model are shown in  Table~\ref{Table:apm}.
Fig.~\ref{fig:apmconstr} represents the marginalized constraints on anapole moment and mass of 
the dark matter along with joint 68\% CL and 95\% CL constraints on both the parameters 
from Planck-2018 and BAO data. 

\begin{table}[h]
\begin{tabular} { l  c c c}
Parameter &  68\% limits & 95\% limits & 99\% limits \\
\hline
{\boldmath$\Omega_b h^2$} & $0.02243\pm 0.00013$ & $0.02243^{+0.00026}_{-0.00026}$
& $0.02243^{+0.00034}_{-0.00034}$\\

{\boldmath$\Omega_c h^2$} & $0.11917\pm 0.00090$ & $0.1192^{+0.0018}_{-0.0018}$
& $0.1192^{+0.0024}_{-0.0023}$\\

{\boldmath$\tau$} & $0.0567\pm 0.0072$ & $0.057^{+0.015}_{-0.014}$
&$0.057^{+0.020}_{-0.018}$ \\

{\boldmath${\rm{ln}}(10^{9}(g_{A}/GeV^{-2}))$} & $< 22.6$ & $<29.6 $ 
 & $< 31.9$\\

{\boldmath${\rm{log_{10}}}(m_{\chi}/GeV)$} & $ --- $ & $ ---$ 
& $ ---$\\

{\boldmath${\rm{ln}}(10^{10} A_s)$} & $3.048\pm 0.014$& $3.048^{+0.029}_{-0.028}$
& $3.048^{+0.039}_{-0.036}   $\\

{\boldmath$n_s            $} & $0.9670\pm 0.0037          $& $0.9670^{+0.0073}_{-0.0073}$
& $0.9670^{+0.0097}_{-0.0096}$\\

$H_0                       $ & $67.73\pm 0.41$ & $67.73^{+0.81}_{-0.80}$
& $67.7^{+1.1}_{-1.1}$\\
\hline
\end{tabular}
\caption{Planck-2018 and BAO constraints on anapole momentum and mass of the dark matter
with other 6 parameters of $\Lambda$CDM}
\label{Table:apm}
\end{table}

It can be seen from Table~\ref{Table:apm} that
\be
g_{A}<7.163\times 10^{3} \text{GeV\textsuperscript{-2}}\;~ 95\% C.L.\label{apmlimit}
\ee

\begin{figure}[h]
\subfigure[Marginalized constraints]{
 \includegraphics[width=7cm, height = 6cm]{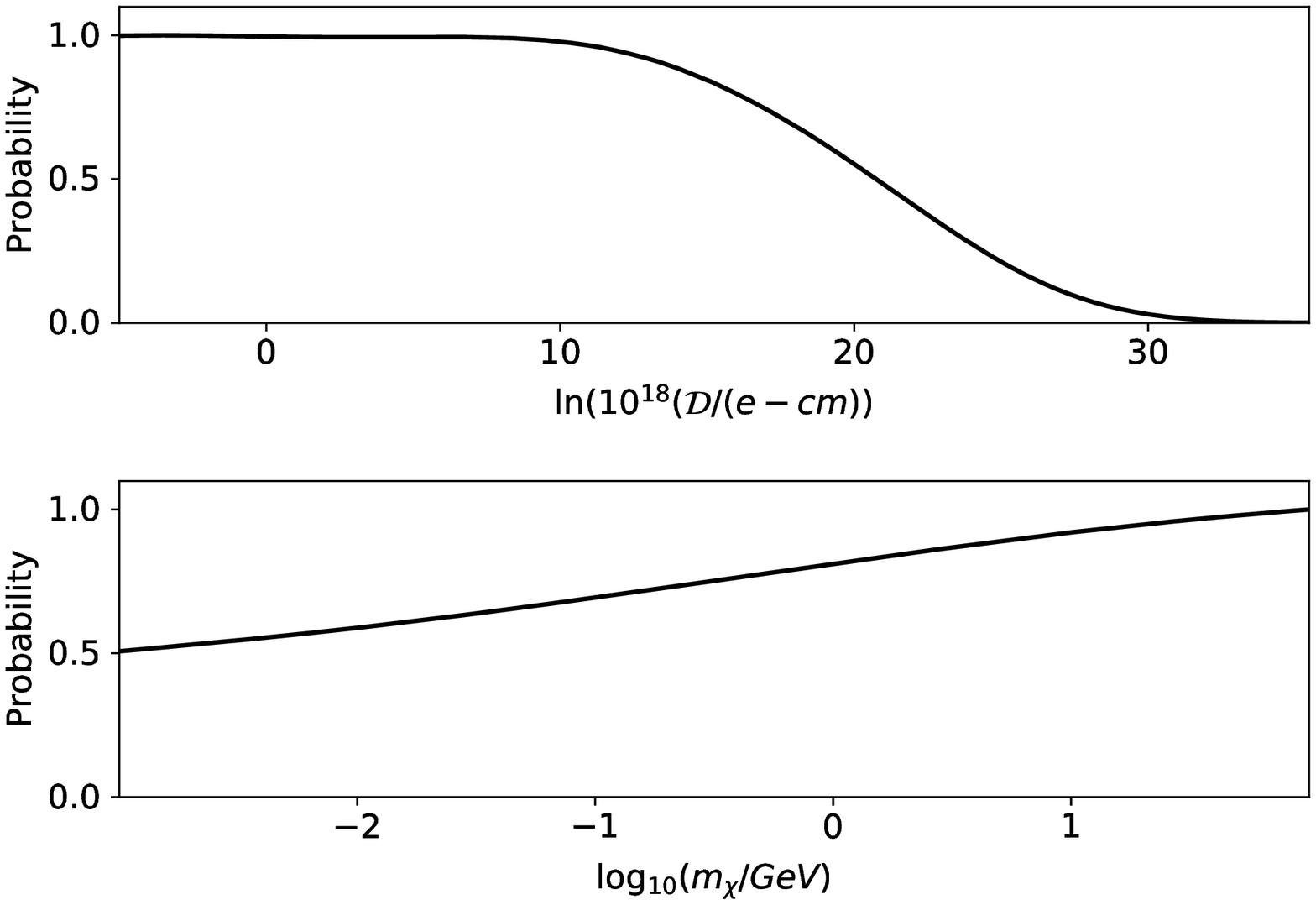}
}
\subfigure[Joint 68\% CL and 95\%CL constraints]{
 \includegraphics[width=7cm, height = 6cm]{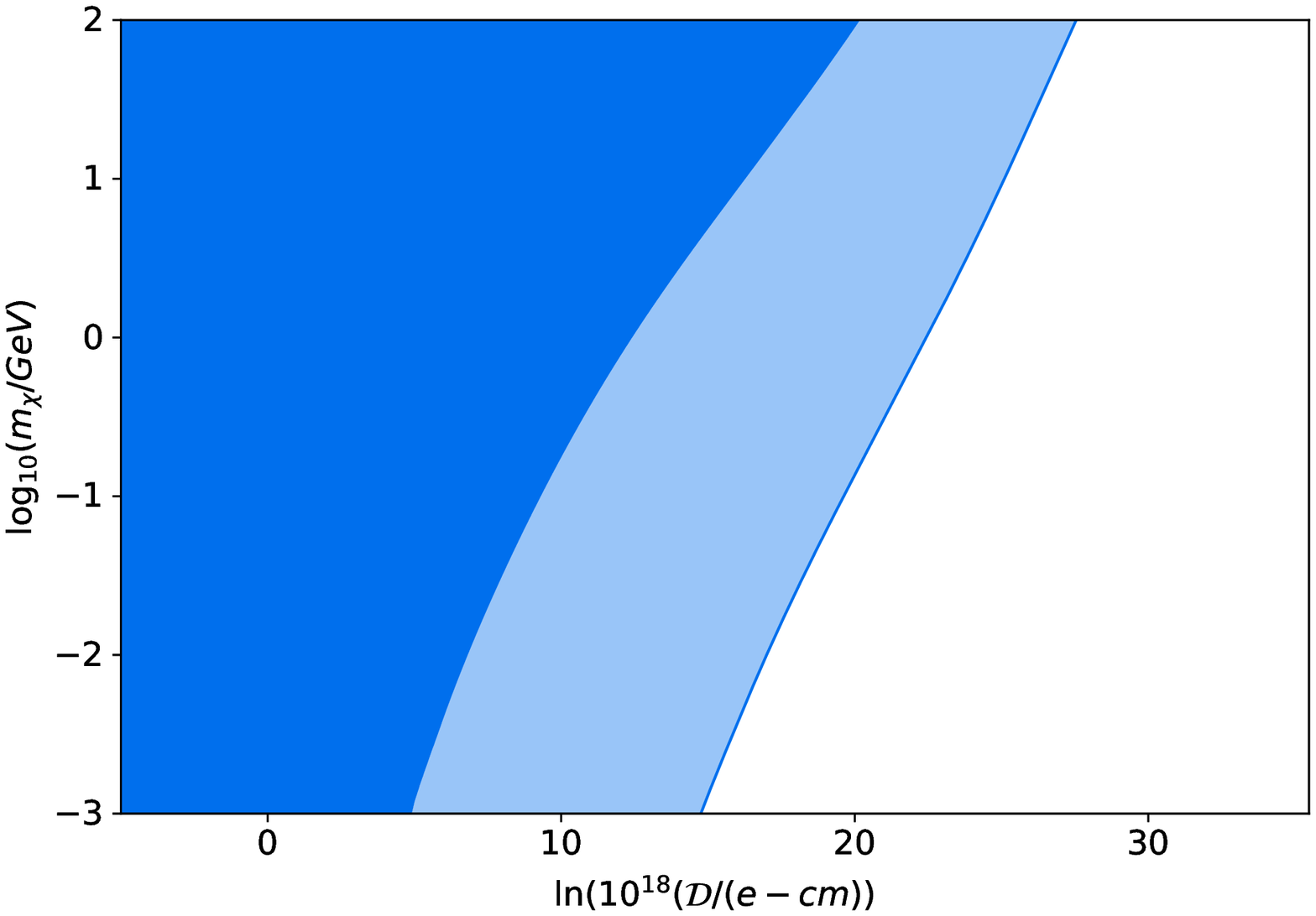}
}
\caption{ Constraints for electric dipole moment and mass of the dark matter
using Planck-2018 and BAO data.}
\label{fig:edmconstr}
\end{figure}

The constraints obtained using Planck-2018 and BAO data on electric dipole moment and the 
mass of the dark matter along with the other six parameters of $\Lambda$CDM model are listed 
in Table.~\ref{table:edm}.
 Fig.~\ref{fig:edmconstr} depicts the marginalized constraints on electric dipole moment and 
mass of the dark matter and with joint 68\% CL and 95\%CL constraints on both the parameters
from Planck-2018 and BAO data.

\begin{table}[h!]
\begin{center}
\begin{tabular} { l  c c c}

 Parameter &  68\% limits  &  95\% limits  &  99\% limits\\
\hline
{\boldmath$\Omega_b h^2   $} & $0.02244\pm 0.00014$ & $0.02244^{+0.00027}_{-0.00026}$
& $0.02244^{+0.00036}_{-0.00034}$\\

{\boldmath$\Omega_c h^2   $} & $0.11921\pm 0.00091$ & $0.1192^{+0.0018}_{-0.0018}$
& $0.1192^{+0.0024}_{-0.0024}$\\
{\boldmath$\tau           $} & $0.0565\pm 0.0073$ & $0.056^{+0.015}_{-0.014}   $
& $0.056^{+0.020}_{-0.018}$\\

{\boldmath${\rm{ln}}(10^{18}({\cal D}/(e-cm))$} & $< 12.9 $  & $< 22.8$
& $< 26.7$\\

{\boldmath${\rm{log_{10}}}(m_{\chi}/GeV)$} & $> -0.963 $  & $---$
& $---$\\

{\boldmath${\rm{ln}}(10^{10} A_s)$} & $3.048 \pm 0.014$ & $3.048^{+0.029}_{-0.028}$
& $3.048^{+0.039}_{-0.037}   $   \\

{\boldmath$n_s            $} & $0.9670\pm 0.0037$ & $0.9670^{+0.0071}_{-0.0073}$
& $0.9670^{+0.0093}_{-0.0097}$\\

$H_0                       $ & $67.72\pm 0.41$ & $67.72^{+0.82}_{-0.79}     $
& $67.7^{+1.1}_{-1.0}$\\

\hline
\end{tabular}
\caption{Planck-2018 and BAO constraints on electric dipole momentum and mass of the dark matter
with other 6 parameters of $\Lambda$CDM}
\label{table:edm}
\end{center}
\end{table}

We can see from Table.~\ref{table:edm} that
\be
{\cal D}<7.978\times 10^{-9} \text{e-cm}\;~ 95\% C.L.\label{edmlimit}
\ee

\begin{figure}[h]
\subfigure[Marginalized constraints]{
 \includegraphics[width=7cm, height = 6cm]{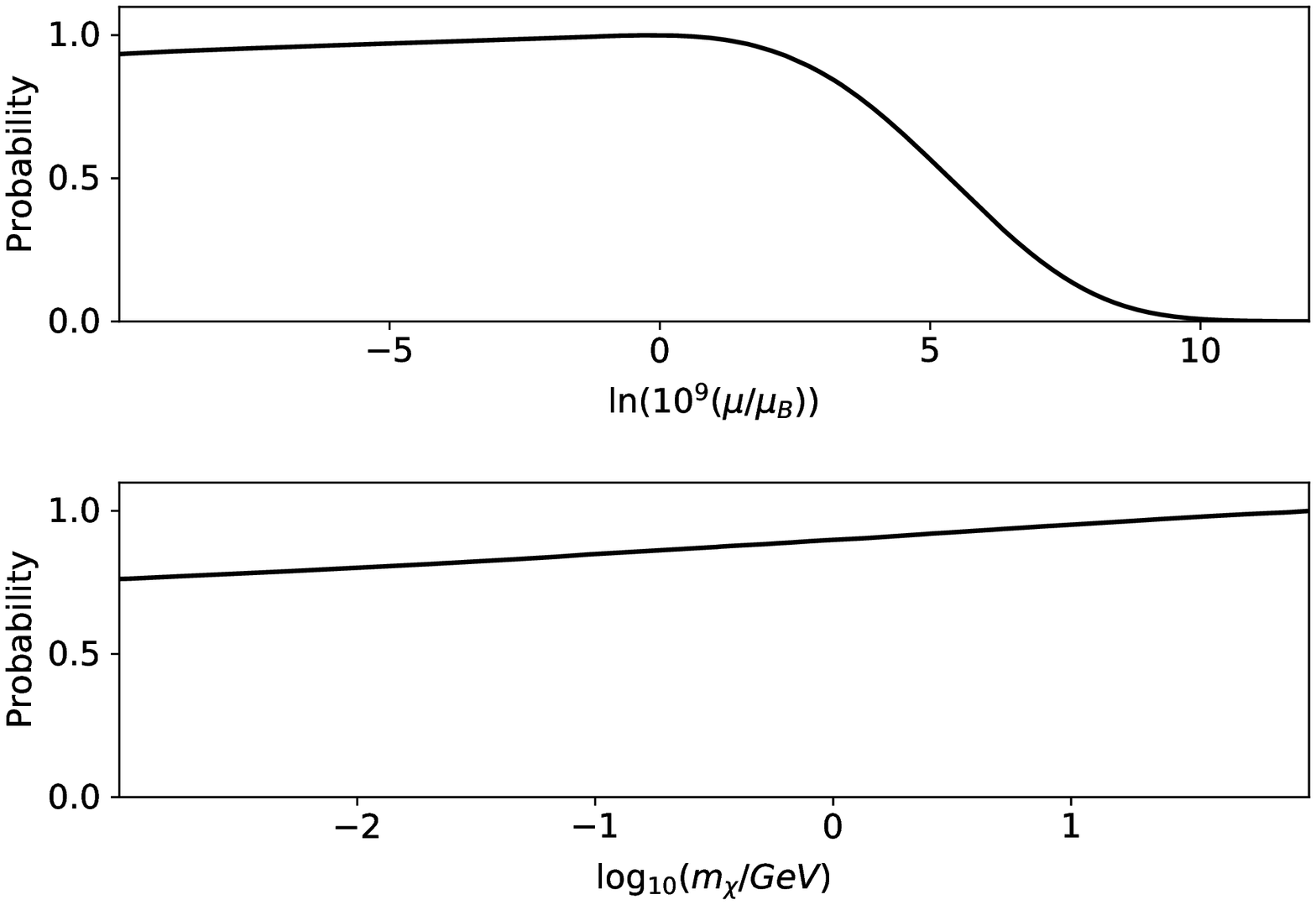}
 \label{fig:mdm1dfig}
}
\subfigure[Joint 68\% CL and 95\%CL constraints]{
 \includegraphics[width=7cm, height = 6cm]{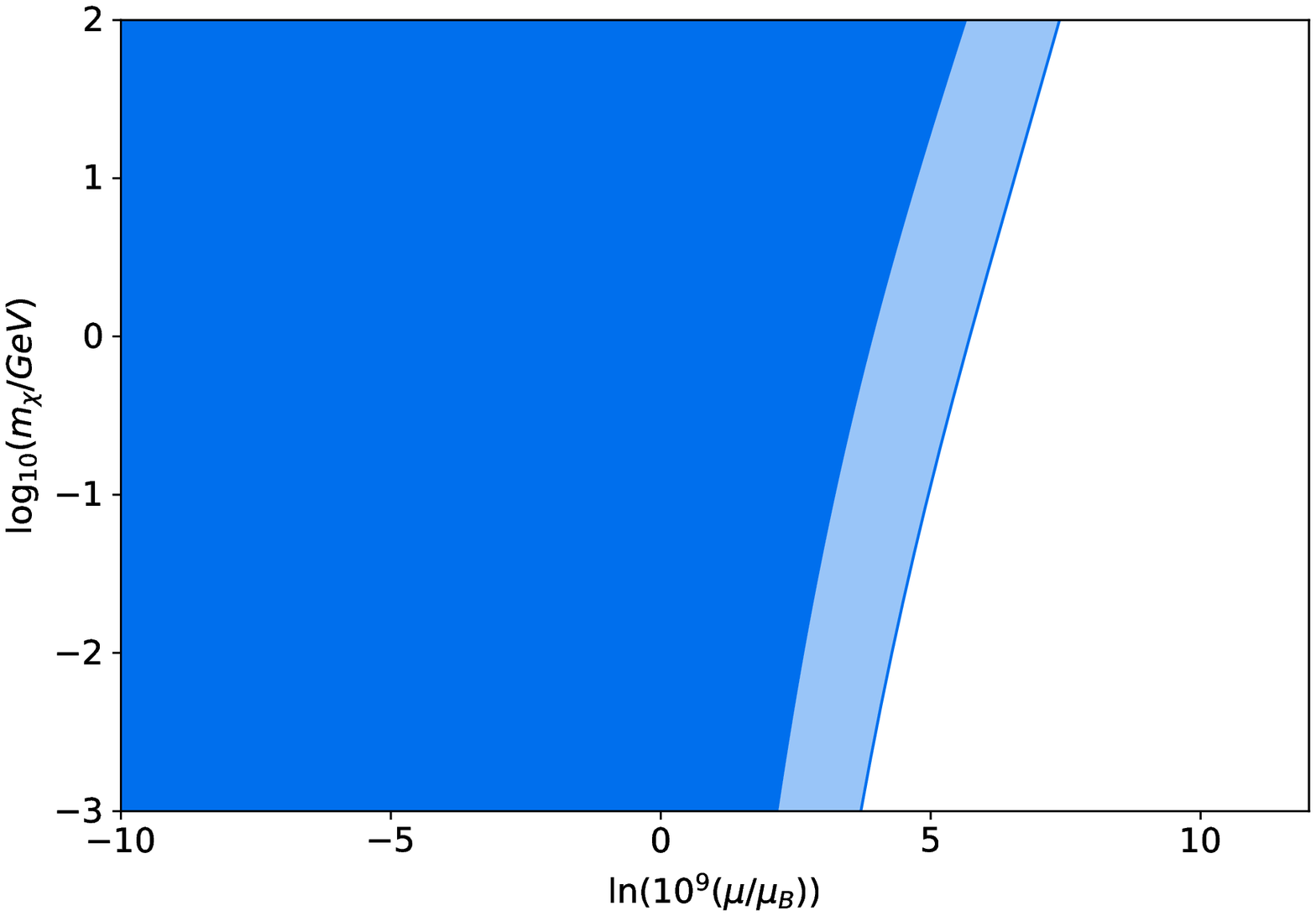}
 \label{fig:mdmmfig}
}
\caption{Constraints for magnetic dipole moment and mass of the dark matter using Planck-2018 and BAO data.}
\label{fig:mdmconstr}
\end{figure}

Table.~\ref{table:mdm} represents the constraints on magnetic dipole moment and mass of the 
dark matter obtained from Planck-2018 and BAO data. Here also we have quoted the constraints
on other six parameters of $\Lambda$CDM. Fig.~\ref{fig:mdmconstr} represents the 
marginalized constraints on magnetic dipole moment and mass of the dark matter along with
joint 68\% CL and 95\%CL constraints on both the parameters. 
%

\begin{table}[h!]
\begin{center}
\begin{tabular} { l  c c c}

 Parameter &  68\% limits  &  95\% limits &  99\% limits\\
\hline
{\boldmath$\Omega_b h^2   $} & $0.02244\pm 0.00013$ & $0.02244^{+0.00026}_{-0.00026}$
& $0.02244^{+0.00034}_{-0.00034}$\\

{\boldmath$\Omega_c h^2   $} & $0.11920\pm 0.00091$& $0.1192^{+0.0018}_{-0.0018}$
& $0.1192^{+0.0023}_{-0.0023}$\\

{\boldmath$\tau           $} & $0.0564 \pm 0.0073$& $0.056^{+0.015}_{-0.014}   $
& $0.056^{+0.020}_{-0.018}   $\\

{\boldmath${\rm{ln}}(10^{9}(\mu/\mu_B))$} & $-2.0^{+4.0}_{-6.4}$ & $< 5.69$ & $< 7.36  $\\

{\boldmath${\rm{log_{10}}}(m_{\chi}/GeV)$} & $ --- $ &$ ---$                   
& $ ....                   $\\

{\boldmath${\rm{ln}}(10^{10} A_s)$} & $3.047\pm 0.014$ & $3.047^{+0.029}_{-0.028}   $
& $3.047^{+0.039}_{-0.037}   $\\

{\boldmath$n_s            $} & $0.9671\pm 0.0037$ & $0.9671^{+0.0073}_{-0.0072}$
& $0.9671^{+0.0096}_{-0.0093}$\\

$H_0                       $ & $67.72\pm 0.41$ & $67.72^{+0.82}_{-0.79}     $
& $67.7^{+1.1}_{-1.0}        $\\
\hline
\end{tabular}
\caption{Planck-2018 and BAO constraints on magnetic dipole momentum and mass of Majorana fermion  dark matter
with other 6 parameters of $\Lambda$CDM}
\label{table:mdm}
\end{center}
\end{table}

Again we can read from Table~\ref{table:mdm}  that

\be
{\mu}<2.959\times 10^{-7} \mu_B\;~ 95\% C.L.\label{mdmlimit}
\ee

Similarly Table.~\ref{table:mdm} lists the constraints on Rayleigh coupling and mass of 
the scalar dark matter.  Fig.~\ref{fig:mdmconstr} represents the marginalized 
constraints on both of these parameters along with joint 68\% CL and 95\%CL constraints from
Planck-2018 and BAO data. 
Again we have mentioned constraints on other six
parameters of $\Lambda$CDM.

\begin{table}[h!]
\begin{center}
\begin{tabular} { l  c c c}

 Parameter &  68\% limits  &  95\% limits &  99\% limits\\
\hline
{\boldmath$\Omega_b h^2   $} & $0.02244\pm 0.00013$  & $0.02244^{+0.00026}_{-0.00026}$
& $0.02244^{+0.00034}_{-0.00034}$\\
{\boldmath$\Omega_c h^2   $} & $0.11919\pm 0.00092$  & $0.1192^{+0.0018}_{-0.0018}$
& $0.1192^{+0.0023}_{-0.0024}$\\

{\boldmath$\tau           $} & $0.0567\pm 0.0074$    & $0.057^{+0.015}_{-0.014}$
& $0.057^{+0.020}_{-0.019}$\\

{\boldmath${\rm{ln}}(10^{9}g_{4}/(\Lambda_{4}^{2} GeV^{-2}))$} & $< 6.43$
& $< 16.2$ & $< 19.7$\\

{\boldmath${\rm{log_{10}}}(m_{\chi}/GeV)$} & $<-3.55$  & --- & ---\\                       \\

{\boldmath${\rm{ln}}(10^{10} A_s)$} & $3.048\pm 0.015$ &$3.048^{+0.030}_{-0.028}$
& $3.048^{+0.040}_{-0.037}$\\

{\boldmath$n_s$} & $0.9671\pm 0.0037$ & $0.9671^{+0.0073}_{-0.0073}$ 
& $0.9671^{+0.0097}_{-0.0096}$\\

$H_0$ & $67.73\pm 0.41$ &$67.73^{+0.81}_{-0.80}$ \\
& $67.7^{+1.1}_{-1.1}$\\
\hline
\end{tabular}
\caption{Planck-2018 and BAO constraints on Rayleigh coupling and mass of
dark matter with other 6 parameters of $\Lambda$CDM}
\label{table:rdm}
\end{center}
\end{table}

\begin{figure}[h]
\begin{center}
\subfigure[Marginalized constraints]{
 \includegraphics[width=7cm, height = 6cm]{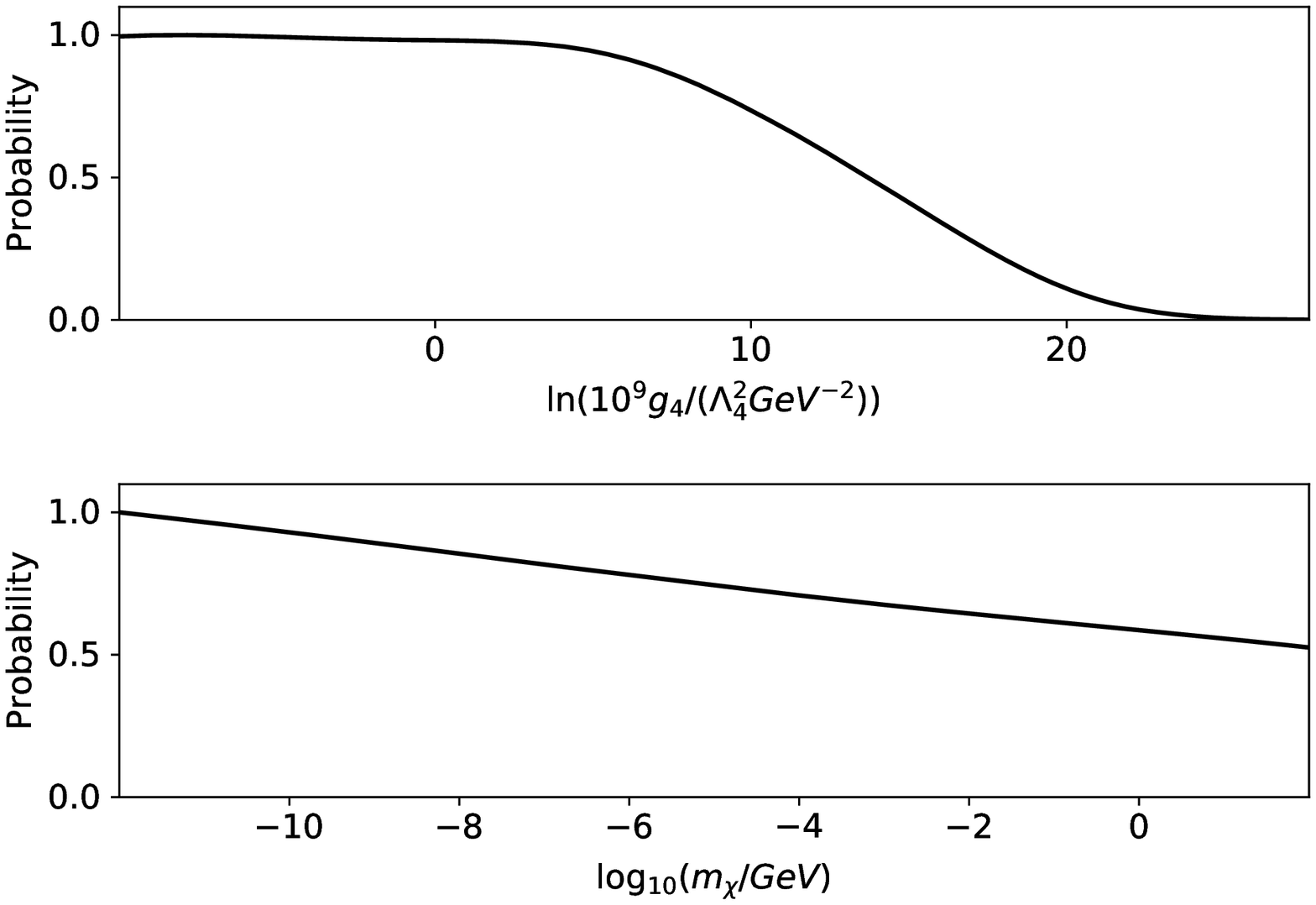}
 \label{fig:rdm1dfig}
}
\subfigure[Joint 68\% CL and 95\%CL constraints]{
 \includegraphics[width=7cm, height = 6cm]{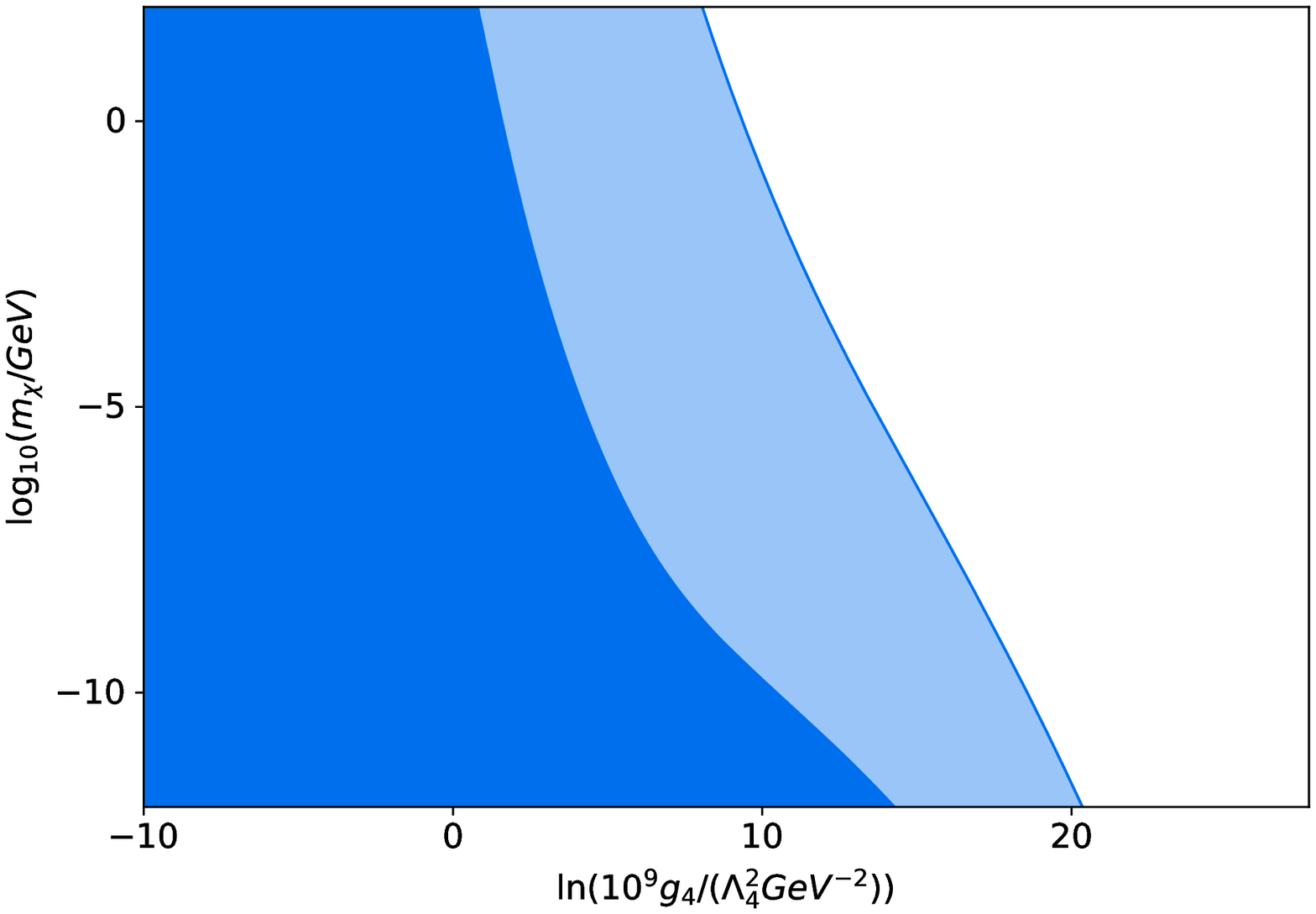}
 \label{fig:rdmmfig}
}
\caption{Constraints on  Rayleigh coupling and mass of scalar dark matter using Planck-2018 and BAO data}
\label{fig:rdmconstr}
\end{center}
\end{figure}

We can see from the Table~\ref{table:rdm} that
\be
{\frac{g_4}{\Lambda_4^2}}<1.085\times 10^{-2}\; \text{GeV\textsuperscript{-2}}\;~ 95\% C.L.\label{rdmlimit}
\ee
The upper bounds given by Eqns.~(\ref{apmlimit}), (\ref{edmlimit}), (\ref{mdmlimit}) 
and (\ref{rdmlimit})  are obtained after margenalizing over all other parameters.

\section{Comparison with constraints from other experiments}\label{Comparison}
In order to obtain the correct relic density of sub-GeV dark matter we require mediators in the sub-GeV mas range \cite{Mohanty:2020pfa}. We can describe dark matter interactions, 
where the energy transfer is so small, with effective operators like anapole, dipole and 
Rayleigh form factors, which are effective below the cutoff scale   $\Lambda \sim {\rm GeV}$.  
These low energy effective form factors cannot be constrained from colliders, and the best 
bounds are obtained from the low energy processes like electron scattering in direct detection 
experiments or in searches of sub-GeV scale gamma rays from galaxies. The constraints which can be obtained on ${\rm TeV}$ scale effective theories from colliders is studied in \cite{Arina:2020mxo}. We compare the bounds from CMB distortion by dark matter annihilation obtained in this paper with the bounds on the electromagnetic form factors from other experiments and astrophysical observations in this section. 

\subsection{Electric, Magnetic and Anapole moments}
For sub-GeV mass  fermionic dark matter the best bounds come from the ionization of atoms with electron scattering process $\chi+e^-\rightarrow \chi +e^-$  in direct detection experiments like Xenon-10 \cite{Essig:2012yx},  DarkSide \cite{Agnes:2018oej} and Xenon-1T \cite{Aprile:2019xxb}. The most stringent bounds come from Xenon-10 \cite{Essig:2012yx} and Xenon-1T \cite{Aprile:2019xxb} which are dark matter mass dependent. 
The bound on anapole moment of Majorana dark matter is $g_A < (10^2 -0.5 \times 10^{-1}) {\rm GeV^{-2}}$ in the mass range $m_\chi=( 0.2 \,{\rm MeV}-1\, {\rm GeV})$ \cite{Catena:2019gfa}. This is more stringent than the CMB bound $g_A < 7.163 \times10^{3}\, {\rm Gev^{-2}}$ (for  $m_\chi \geq 0.5 \, {\rm MeV}$) since the CMB bound is based on annihilation process  $\chi \chi \rightarrow e^+ e^-$, whose cross section is velocity suppressed (\ref{sigmaann}).

The  bound on electric dipole moment of Xenon-10 \cite{Essig:2012yx} and Xenon-1T \cite{Aprile:2019xxb} of Dirac dark matter with mass in the range $m_\chi=( 0.2 \,{\rm MeV}-1\, {\rm GeV})$ is ${\cal D} < (6.6 \times 10^{-18} -3.9 \times 10^{-20})\,  {\rm e-cm}$  \cite{Catena:2019gfa}. This is again more  stringent than the CMB bound ${\cal D} <7.978  \times 10^{-9}\, \text{ e-cm}$ (for $m_\chi \geq 0.5 {\rm MeV}$) as  the annihilation cross section for electric dipole annihilation is velocity suppressed  (\ref{Dann}).

Finally  bound on magnetic dipole moment from Xenon-10 \cite{Essig:2012yx} and Xenon-1T \cite{Aprile:2019xxb} of Dirac dark matter with mass in the range $m_\chi=( 0.2 \,{\rm MeV}-1\, {\rm GeV})$ is $\mu < (1.2 \times 10^{-5} -5.9 \times 10^{-8}) \mu_B$  \cite{Catena:2019gfa}. This is comparable to the CMB bound $\mu < 2.959 \times 10^{-7} \mu_B $ (for $m_\chi \geq 0.5 {\rm MeV}$). The CMB bound is comparable with the direct detection bounds as the annihilation cross section for magnetic dipole annihilation  is not velocity suppressed  (\ref{magann}).

\subsection{Rayleigh form factor of scalar dark matter}
Constraints on the Rayleigh form factor (\ref{Rayleigh}) are obtained from electron ionisation by Xenon-1T \cite{Aprile:2019xxb} and by searches for $\gamma$ ray line spectrum in the Milky Way center by Fermi-LAT \cite{Ackermann:2015lka}. For dark matter of mass $\sim {\rm GeV}$ the bound from Fermi-LAT search is $g_4/\Lambda_4^2 < 10^{-3} \,{\rm GeV^{-2}}$, and from Xenon-1T the bound is $g_4/\Lambda_4^2 < 0.25 \,{\rm GeV^{-2}}$ \cite{kavanagh}. The CMB bound obtained in the paper $g_4^2/\Lambda_4^2 < 1.1\times 10^{-2} \, {\rm GeV^{-2}}$ is weaker than the Fermi-LAT bound but is valid for larger range of dark matter masses up to  $m_\phi \geq {\rm eV}$. 

\section{Conclusions}\label{Conclusions} Electromagnetic form factors are an important class of interactions in the effective theories framework of classifying dark matter interactions. 
 Dirac fermions  dark matter with non-zero electric 
and magnetic  dipole moments can give the correct relic density $\Omega_m h^2 = 0.11$ by the $\chi \chi \leftrightarrow f \bar f$ freeze-out process if ${\cal D} = 2.5 \times 10^{-16}\, 
\text{ e-cm}$ and $\mu=  8.2 \times 10^{-7} \mu_B $ respectively \cite{Masso:2009mu}.  Majorana fermions with anapole moment of mass $10$ MeV with anapole moment $g_A=0.11 {\rm Gev^{-2}}$ can be dark matter with the correct  freeze-out relic density \cite{Ho:2012bg}.

Electromagnetic dark matter can be observed not only via electron scattering direct detection 
experiments \cite{Essig:2011nj,Essig:2012yx, Agnes:2018oej,Aprile:2019xxb, Catena:2019gfa,Catena:2020tbv} but can also be constrained from the CMB. 

In this paper we have considered anapole and dipolar dark matter matter with masses $m_\chi > \cal{O}({\rm MeV})$. We find that  dark matter with electromagnetic dipole or anapole form factors will distort the 
CMB during recombination era by producing relativistic electron via the process 
$\chi \chi \rightarrow e^+ e^-$. We find that the Planck data gives the bounds on 
electromagnetic form factors ${\cal D} <7.978  \times 10^{-9}\, \text{ e-cm}$ and 
$\mu < 2.959 \times 10^{-7} \mu_B $, and $g_A < 7.163 \times10^{3} {\rm Gev^{-2}}$.

Dark matter with $\cal{O}({\rm MeV})$ mass in thermal equilibrium with radiation will be ruled 
out by BBN constraints on $N_{\rm eff}$. These can only have be created after the BBN era by 
the freeze-in mechanism. Freeze-in requires very small couplings and our bounds on electric 
dipole and anapole moments also rules out the freeze-out mechanism for relic density. 
These may be produced by the freeze-in mechanism which requires smaller 
couplings \cite{Hall:2009bx, Mohanty:2020pfa}.   

For scalar dark matter there is the dimension-six Rayleigh operator coupling with photons. We put the bound on the Rayleigh coupling as
$\frac{g_4}{\Lambda_4^2}<1.085\times 10^{-2} \text{GeV\textsuperscript{-2}}$ ( $95 \%$C.L).This bound is valid for dark matter mass as low as $\cal{O}({\rm eV})$. Such light dark matter can only be produced by the freeze-in mechanism to evade  bounds from BBN.

Spectral distortion of the CMB can also arise from radiatively decaying dark matter 
\cite{Bolliet:2020ofj}. The bounds derived on the radiative lifetime can be used for deriving bounds on dipolar couplings of Majorana dark matter which can have non-zero transition electric and  magnetic moments dipole.

\section{ACKNOWLEDGEMENTS}
A. N. would like to thank ISRO Department of Space Govt. of India to provide financial 
support via RESPOND programme Grant No. DS\_2B-13012(2)/47/2018-Sec.II.  

\end{document}